\documentclass[english,aps,prl,twocolumn, superscriptaddress,showpacs, amsmath]{revtex4-1}
\usepackage[T1]{fontenc}
\usepackage[latin9]{inputenc}
\usepackage{color}
\usepackage{units}
\usepackage{amssymb}
\usepackage{graphicx}
\usepackage{esint}
\usepackage{bm}
\usepackage{natbib}

\usepackage[normalem]{ulem}

\usepackage{hyperref}
\hypersetup{colorlinks=true,
	    citecolor=blue,
	    linkcolor=blue,
            urlcolor = blue,	   
}

\usepackage{graphicx}% Include figure files
\usepackage{dcolumn}% Align table columns on decimal point
\usepackage{bm}% bold math

\usepackage{babel}
\usepackage{comment}
\usepackage{soul}

\newcommand{\tc}{$ T_{C} $ }

\begin{document}
%%%%%%%%%%%%%%%%%%%%%%%%%%%%   Title and Author
\title{Universal scaling behaviour near vortex-solid/glass to vortex-fluid transition in type-II superconductors in two- and three-dimensions}

\author{Hemanta Kumar Kundu}
\email{hemantakundu@iisc.ac.in}
\affiliation{Department of Physics, Indian Institute of Science, Bangalore 560012, India}
\author{John Jesudasan}
\affiliation{Tata Institute of Fundamental Research, Mumbai 400005, India}
\author{Pratap Raychaudhuri}
\affiliation{Tata Institute of Fundamental Research, Mumbai 400005, India}
\author{Subroto Mukerjee}
\affiliation{Department of Physics, Indian Institute of Science, Bangalore 560012, India}
\author{Aveek Bid}
\affiliation{Department of Physics, Indian Institute of Science, Bangalore 560012, India}

\begin{abstract}
	
In this article, we present evidence for the existence of vortex-solid/glass (VG) to vortex-fluid (VF) transition in a type-II superconductor (SC), NbN. We probed the VG to VF transition in both 2D and 3D films of NbN through studies of magnetoresistance and current-voltage characteristics. The dynamical exponents corresponding to this phase transition were extracted independently from the two sets of measurements. The $H$-$T$ phase diagram for the 2D and 3D SC are found to be significantly different near the critical point. In the case of 3D SC, the exponent values obtained from the two independent measurements show excellent match.  On the other hand, for the 2D SC, the exponents obtained from the two experiments were significantly different. We attribute this to the fact that the characteristic length scale diverges near the critical point in a 2D SC in a distinctly different way from its 3D counterpart form scaling behaviour. \\

%\noindent {Keywords: superconductivity, electronic transport, vortex lattice, vortex melting}

\end{abstract}

\maketitle
%\section{Introduction}
A fundamental property of a superconductor -- known as the Meissner effect -- is its ability to expel a magnetic field $H$ from its interior~\cite{tinkham2004introduction}. At high magnetic fields, this property is compromised in type-II superconductors and magnetic-flux threads in leading to the formation of topological defects known as vortices. Vortex dynamics lies at the heart of dissipationless transport in type-II superconductors.  Vortices by nature interact with each other repulsively. At a given applied magnetic field, they arrange themselves in an ordered hexagonal  lattice~\cite{abrikosov1957magnetic,blatter1994vortices}. A finite applied current can cause the entire vortex-solid to move due to Lorentz force and create an electric field along the direction of the applied current inducing dissipation in the system~\cite{PhysRev.140.A1197,abrikosov1957magnetic}. Such a system cannot support dissipationless transport even for infinitesimally small electrical currents.

In any real sample however, the ubiquitous disorders  pin the vortices restoring the true zero-resistance state~\cite{PhysRev.178.657,higgins1996varieties}. Pinning destroys the long range positional and orientational order of the vortex-solid -- there is only local ordering in the lattice and the vortex state is called a  `vortex-glass'~\cite{fisher1991thermal,PhysRevB.52.1242,PhysRevB.55.6577,PhysRevB.40.11355}. Both magnetic and electric fields can tune the density of the vortex-solid/glass and affect the dissipation induced in the superconductor~\cite{PhysRev.178.657,PhysRev.140.A1197}. At large enough fields/temperatures  the vortex-solid/glass can melt into a dissipative vortex-fluid state~\cite{PhysRevLett.72.2951,zeldov1995thermodynamic,PhysRevLett.76.835,schilling1996calorimetric,PhysRevLett.78.4833,sun2013voltage}. 

Despite decades of research a clear picture of the solid-fluid transition in 2D superconductors is elusive~\cite{PhysRevB.63.024505,PhysRevB.74.104502,PhysRevLett.86.712,PhysRevLett.85.3712,PhysRevB.62.9191,PhysRevLett.88.167003,PhysRevB.82.144508,PhysRevLett.70.505,PhysRevB.47.262,PhysRevLett.70.505,PhysRevLett.65.2583}. {The effect of dimensionality and disorder on the order and nature of this transition is  widely debated~\cite{blatter1994vortices,PhysRevB.75.184532,PhysRevLett.96.177001,PhysRevB.83.174501,PhysRevB.19.3580,soibel2000imaging,ozer2006hard}.} There have been predictions, and some recent experimental verification   of interesting intermediate phases like hexatic phases~\cite{guillamon2009direct,PhysRevLett.122.047001}  %Recently, there has been extensive discussions in the community regarding the possibility of the existence of 2D superconductors in dissipative state at any arbitrarily-low temperature in the presence of a magnetic field (called  Bose metal state)  
and the absence of  dissipationless state in a single sheet of superconductor because of vortex dynamics~\cite{tsen2016nature,sharma20182d,benyamini2019absence,tamir2019sensitivity}. Vortex-solid/glass is also of immense practical interest because it is an important parameter governing the applications of high-T$_C$ superconductors~\cite{PhysRevLett.63.1511,blatter1994vortices}. High-\tc materials are known to exhibit the Berezinskii-Kosterlitz-Thouless (BKT)  transition which is a hallmark of 2D superconductors~\cite{PhysRevB.42.2242}. Presence of unavoidable topological defects like vortices and anti-vortices  makes it complicated to analyze their zero-resistance state~\cite{kosterlitz2016early,PhysRevLett.70.670,PhysRevB.47.262,PhysRevLett.70.505}. This is partially the reason why clear experimental signatures of the existence of vortex-solid/glass remain elusive so far. 

\begin{figure}[t]
	\begin{center}
		\includegraphics[width=0.48\textwidth]{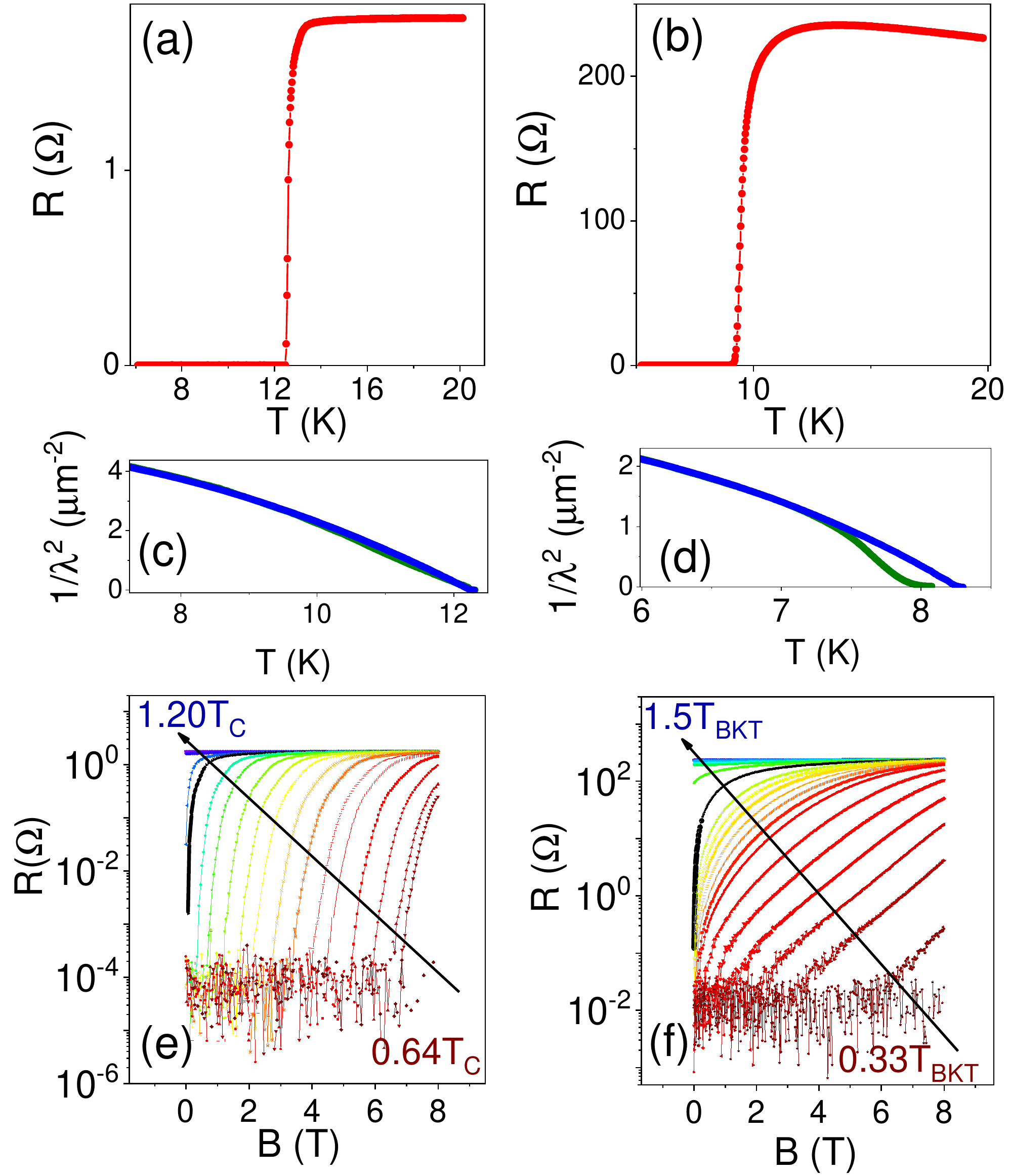}
		\small{\caption{Resistance versus temperature plot for (a) 68~nm NbN film, and (b) for 3~nm NbN film. (c) and (d) are plots of the variation of super-fluid number density $n_s$ versus $T$ for 3D and 2D superconductors respectively. In both cases the green line is the measured data and the blue line is a fit to the BCS relation. In  the case of the 2D film, there is an universal  jump of $n_s$ to zero at $T_{BKT}$.  For the 3D superconductor, the $T$ dependence of $n_s$  follows the BCS prediction over the entire range of $T$. (e) and (f) are magnetoresistance plots for 3D and 2D NbN superconductors respectively at different $T$. The black thick lines show the magnetoresistance data measured at the transition temperature.
				\label{fig:Graph1}}}
	\end{center}
\end{figure} 

%\section{Results}
In this article we report a detailed, comparative study of magnetoresistance and $H$ dependence of current-voltage ($I$-$V$) characteristics  of  high-quality NbN films of two thicknesses -- 3~nm and 68~nm. NbN is a well known, conventional type-II superconductor with excellent control over induced disorder and thickness~\cite{PhysRevB.79.094509,PhysRevLett.106.047001}. {Both 2D and 3D homogeneous NbN films  of high crystalline quality can be grown which exhibit excellent superconducting properties~\cite{PhysRevLett.107.217003}}. It lacks the plethora of phases like high-\tc materials, making it easier to interpret the experiential observations. The coherence length of NbN, as obtained from upper critical field measurements, is $\sim$~6~nm~\cite{PhysRevLett.111.197001}. {Both transport, and super-fluid density studies have established that NbN films of thickness 3~nm undergo BKT transition, making this material an ideal candidate for a comparative study of vortex dynamics in 2D  and 3D  superconductors~\cite{PhysRevLett.107.217003,PhysRevLett.111.197001}.} 

We find that the superconductor to normal transition, in presence of a perpendicular magnetic field, is characteristically different for the 3~nm~(2D) and 68~nm~(3D) superconductors at $T$ close to $T_C$. For $T<<T_C$, the $T$--$H_{C2}$  line becomes similar in nature for the two systems. It is in  this regime that we looked for the possible existence of vortex-solid/glass to fluid transition in a 2D superconducting film and  compared its characteristics to the well-documented corresponding transition of a 3D superconductor. 

The films were grown by reactive dc magnetron sputtering of a Nb target in Ar-N$_2$ gas mixture~\cite{PhysRevB.77.214503,PhysRevB.80.134514} and were characterized by measurements of temperature dependence of resistance, dc $I$-$V$ characteristics, magnetoresistance and super-fluid density. Figure~\ref{fig:Graph1}(a) shows the resistance $R$ \textit{versus} temperature $T$ data for the 68~nm film. For this 3D film, the superconducting transition temperature \tc (defined throughout in this article as the temperature where  $R$ drops to 1\% of its normal state value)  is 12.44~K, very close to that of bulk single crystal NbN. In Fig.~\ref{fig:Graph1}(c) is plotted the measured variation of super-fluid density (which is proportional to the inverse square of the penetration depth, $\lambda^{-2}$) with temperature.  The data follows the classic BCS pattern as $T$ is increased from deep inside superconducting regime and goes to zero at \tc as expected. The corresponding data for the 3~nm film are shown in Fig.~\ref{fig:Graph1}(b) and Fig.~\ref{fig:Graph1}(d).  The mean field transition temperature $T_C$ (T where R drops to 1\% of its normal state value) is 9.2~K and the  $T_{BKT}$, identified by the universal jump in $\lambda^{-2}$ (or equivalently, the super-fluid density). %\textcolor{red}{\sout{, is 9.07~K.}} 
A plot of $\lambda^{-2}$ versus $T$ is shown in Fig.~\ref{fig:Graph1}(d)~\cite{Nam10513}.  Note that due to the differing requirements of sample geometry and sizes for the transport and $\lambda^{-2}$ measurements, these were measured on two different NbN films, both of thickness 3~nm and hence the slightly different $T_C$ and $T_{BKT}$. The identification of $T_{BKT}$ on the measured device is obtained by power law behaviour of current-voltage (I-V) characteristics in superconducting state~\cite{tsen2016nature,RevModPhys.59.1001,PhysRevB.39.9708,PhysRevB.42.2242,PhysRevB.94.085104,Reyren1196}.

Figure~\ref{fig:Graph1}(e) and (f) show the magnetoresistance data for 3D and 2D superconducting films respectively at different $T$ for $H$ applied perpendicular to the plane of the films. We observe a striking difference in the way  dissipation appears in these two cases. In  3D film, a finite amount of $H$ is necessary to destroy superconductivity at all temperatures. In contrast, for 2D superconductor near $T_C$, an infinitesimal magnitude of $H$ is enough to destroy superconductivity. We find that at $T/T_C =0.9$ one needs $H=2.68$~T to induce dissipation in the 3D case as opposed to only 0.22~T in the 2D case -  an order of magnitude lesser than the 3D case. This significant difference in magnetoresistance can be appreciated better from the surface plots shown in Fig.~\ref{fig:Graph2}(a) and Fig.~\ref{fig:Graph2}(b) for 3D and 2D respectively. The color bar represents the resistance in logarithmic scale normalized by its normal state value. One can  clearly see that there is a distinct difference in the way superconductivity is destroyed near \tc (marked in Fig.~\ref{fig:Graph2} by a black dotted circle) in presence of a perpendicular magnetic field. {We have measured three different NbN films of thickness 3~nm -- the data in all of them were qualitatively similar. Here we present data of one representative sample where we have studied the phase diagram extensively.}

\begin{figure}
	\begin{center}
		\includegraphics[width=0.48\textwidth]{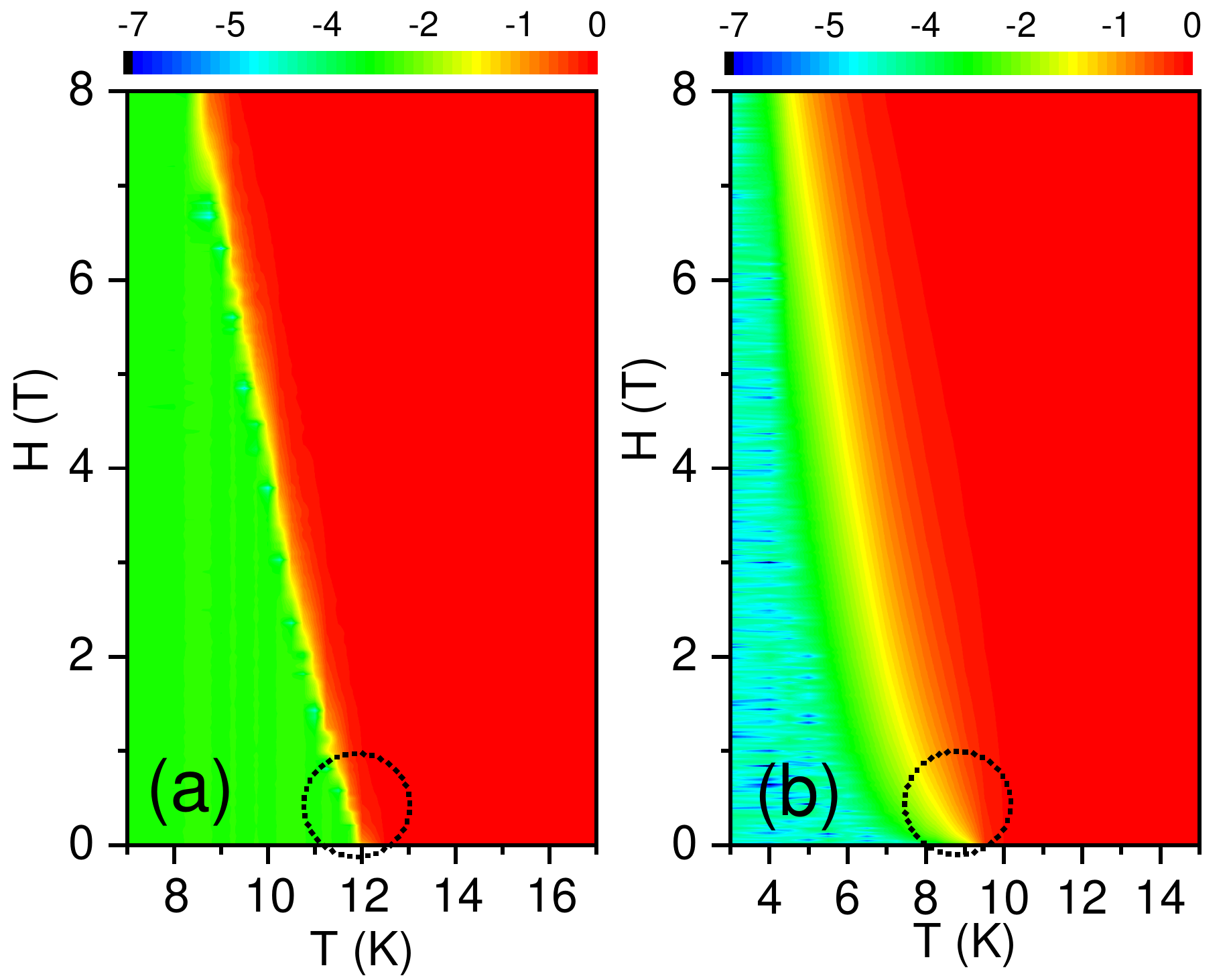}
		\small{\caption{(a) and (b) surface plots of log$({R}/{R_N})$ versus $T$ and $H$ for the 3D and 2D NbN films respectively. The dotted black circles highlight the transition region. It can be seen that close to $T_C$, in the 2D superconducting film the infinitesimal amount of perpendicular field is enough to cause dissipative transport whereas in the 3D film a finite magnetic field is required over the entire temperature range to get some dissipation. 
				\label{fig:Graph2}}}
	\end{center}
\end{figure}

To look into these differences quantitatively, we plot in fig.~\ref{fig:Graph3}(a) the critical magnetic field ($H_C$) versus temperature for both 3D (red solid line) and 2D (black solid line) films. $H_C$  is defined at the field at which  $R(H)$  reaches 1\% of its normal state resistance at each corresponding temperature. Near $T_c$, the two curves have very different characteristics becoming quite similar at lower temperatures. 

\begin{figure}[t]
	\begin{center}
		\includegraphics[width=0.48\textwidth]{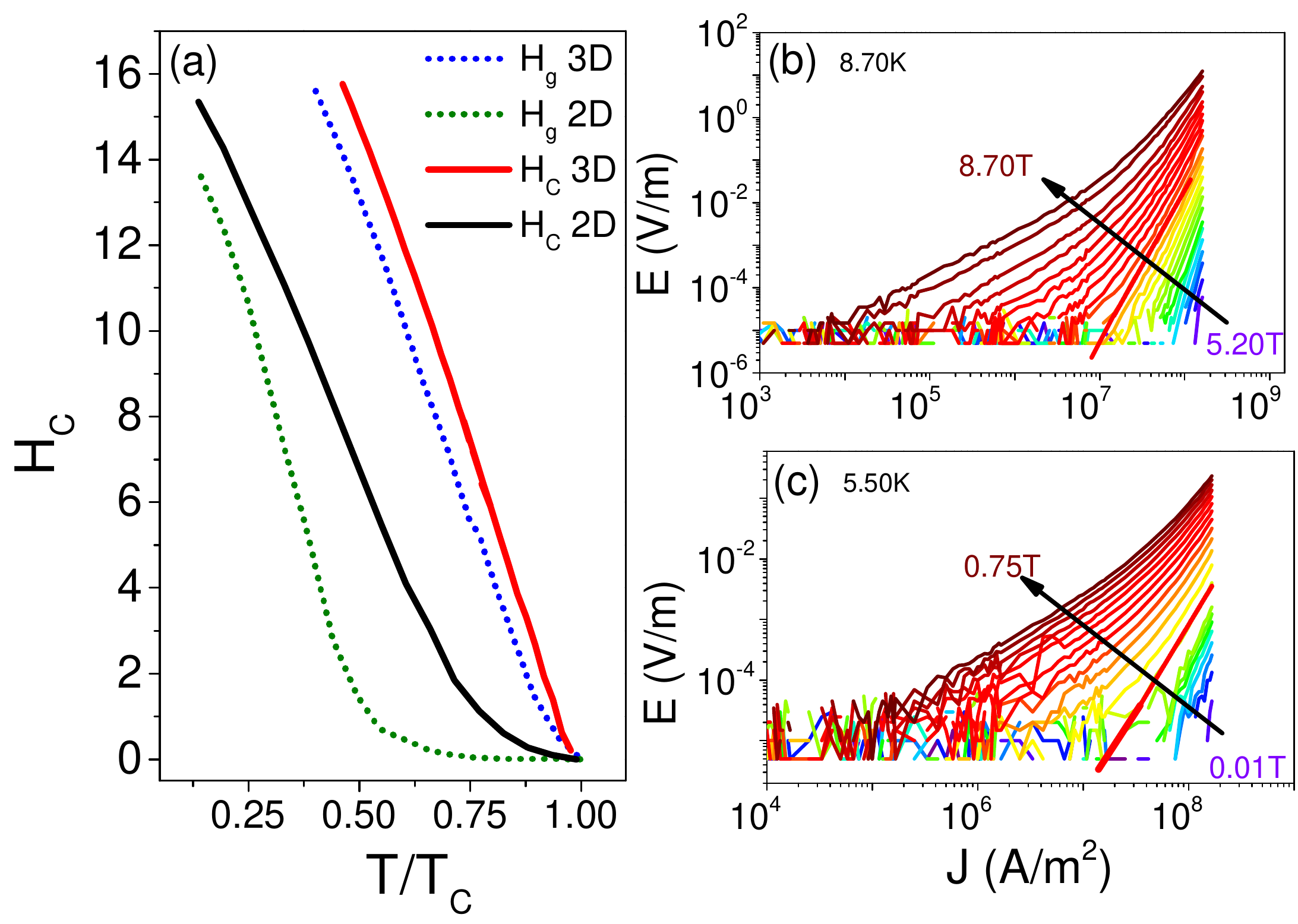}
		\small{\caption{(a) Plot showing $H_C$ versus $T/T_C$ for both the 2D (black solid line) and 3D superconductors (red solid line).  Green and blue dotted lines show the plot of $H_g$ (the field at which the transition takes place from vortex-solid/glass to vortex-fluid) versus $T/T_C$ for the 2D and 3D superconductors, respectively. (b) and (c): representative dc $E$-$J$ characteristics  for 3D film at 8.7~K and for 2D film at 5.5~K, respectively. The $E$-$J$ curves were obtained at different magnetic fields at intervals of 0.20~T.
				\label{fig:Graph3}}}
	\end{center}
\end{figure}
%\section{Discussion} 
A possible reason for this very striking difference between the response of 2D and 3D to  $H$ close to \tc can be the presence of a vortex-solid/glass at low temperatures which undergoes a phase transition to a vortex-fluid in a manner different  in the 3D than in the 2D superconductor  in the presence of the field. Near a critical point of a  phase transition, physical quantities can be expressed in terms of correlation length, $\xi$ and time, $\xi^z$ where $z$ is the dynamical exponent. The critical point is temperature and magnetic field dependent. $H_g(T)$ is the field at which this vortex-solid/glass to vortex-fluid transition occurs at a given $T$. Since the critical line exists in the $H-T$ plane, the correlation length $\xi$ can diverge as this line is approached by changing the magnetic field at constant temperature. We assume that $\xi$ diverges with an exponent $\alpha$, $\xi \sim \lvert H-H_g\rvert^{-\alpha}$ near the vortex-solid/glass to vortex-fluid transition. While, the exponent $\alpha$ might in general be expected to equal the correlation length exponent obtained at constant field by changing the temperature, we do not explicitly require a consideration of this equality since in our measurements, we always approach the transition by changing the field. 

For vortex-solid/glass to vortex-fluid transition, the dc $E$-$J$ characteristic for a d-dimensional superconductor should follow the following scaling relation~\cite{fisher1991thermal,PhysRevLett.63.1511}:
\begin{equation}\label{mastereq}
E(J) \approx J\xi^{d-2-z}\tilde{E}_\pm(J\xi^{d-1}\phi_0/K_BT)
\end{equation}
where $J$ is the current density and $E$ is the electric field. In the scaling function $\tilde{E_\pm} (x)$, $J$ is scaled by the characteristic current density $J_0 ={ K_B T}/(\phi_0 \xi ^{d-1})$ which vanishes as $H \rightarrow H_g$ from below. Under the condition $H \rightarrow H_g$ from above and $J\rightarrow0$,  $\tilde{E_+}(J/J_0)$ becomes a constant giving a resistivity $\rho_L$ which vanishes as:
\begin{equation}\label{bgmr}
\rho_L(H) \sim (H-H_g)^{\alpha(z+2-d)}
\end{equation}
Eqn.~\ref{mastereq} indicates that for $J/J_0\rightarrow\infty$, $\tilde{E}_{\pm}\sim x^{z+2-d)/(d-1)}$. This implies that at $H = H_g$, the $E$-$J$ characteristics should follow the power-law
\begin{equation}\label{bgiv}
E(J;H=H_g) \approx J^{(z+1)/(d-1)}
\end{equation}
We measured the $E$-$J$ characteristics of both the 2D and 3D NbN films  over a range of  values of $T$ and $H$. From Eqn.~\ref{bgiv} we expect that, at a given $T$,  the $E$-$J$ curves for large enough currents should be a power law  for $H \sim H_g$. At low currents, the behavior should be ohmic for $H>H_g$ and exponentially vanishing dissipation for $H<H_g$.

Figure~\ref{fig:Graph3} (b) and (c) are the plots of $E$-$J$ characteristics at a few representative values of $H$ for the 3D (measured at 8.7~K) and for  2D (measured at 5.5~K) superconducting films respectively. Similar data have been presented in the supplementary material for several different temperatures for both the 3D and 2D film. %\textcolor{blue}{Note that we do not observe any indication of Joule heating in our operating current density range.} 
At each temperature, we find that the $E$-$J$ curves show a power-law behavior for a particular value of $H$, which we identify to be  $H_g(T)$.  For $H>H_g$, the $E$-$J$ curve have a  positive curvature and at low $J$ there is an ohmic regime described by linear resistivity $\rho_L\neq 0$. For $H<H_g$, the $E$-$J$ curves have a negative curvature with $E$ going to zero rapidly as $J$ is decreased. At $H=H_g$, one can fit the $E$-$J$ curves to a power law and extract the  dynamical exponent $z$, as given by Eqn.~\ref{bgiv}. The power law fits are shown in red solid line in fig.~\ref{fig:Graph3}(b) and \ref{fig:Graph3}(c).  The values of $z$ obtained at different $T$ are listed in table~\ref{2d3dtable}. Note that all these fits were performed in the very low current density limit ie. $J<<J_{FF}$ (where $J_{FF}$ is the flux-flow current density) or equivalently, for $R << R_{N}\frac{H}{H_{C2}(0)}$~\cite{PhysRev.139.A1163}. In Fig.~\ref{fig:Graph3}(a) the $H_g$  lines are plotted versus $T$  for both the 2D and 3D superconductors. We observe for 3D, the $H_C$  line and $H_g$  line do not differ much whereas there is significant distinction for the 2D superconductor.

The critical current density, $J_0$ was obtained from the $E$-$J$ plots. For $H>H_g$ and  $J<J_0$, the $E$-$J$ characteristics are linear. For  $J>J_0$ it follows an asymptotic behaviour. $J_0$ can be thus be determined by noting the $J$ value beyond which the $E$-$J$ characteristics deviate significantly  from linearity  for H>$H_g$. We used  the criterion  $\frac{\delta logE}{\delta logJ} = 2$ to define $J_0$. The  value 2 is not special, other choices of slope of the log($E$) versus log($J$) plots between 1.5  to 3, while changing the value of $J_0$,  leave the scaling behavior unchanged. $J_0$ has a power law dependence  on ($H-H_g$): 
\begin{eqnarray}
J_0 = \frac{ k_B T}{\phi_0 \xi ^{d-1}} =  \frac{ k_B T}{\phi_0} (H-H_g)^{\alpha{(d-1)}} \label{eqn:j0}
\end{eqnarray} 
From the power-law fits to plots of $J_0$ versus $(H-H_g)/H_g$  we obtained the value of exponent $\alpha$. Figures~\ref{fig:J0_R}(a) and ~\ref{fig:J0_R}(b) show the plots of $J_0$ versus $(H-H_g)/H_g$ obtained at representative temperatures for the 3D and 2D superconductors respectively. The values of $\alpha$ extracted from these plots are tabulated in table~\ref{2d3dtable}.

\begin{figure}
	\begin{center}
		\includegraphics[width=0.48\textwidth]{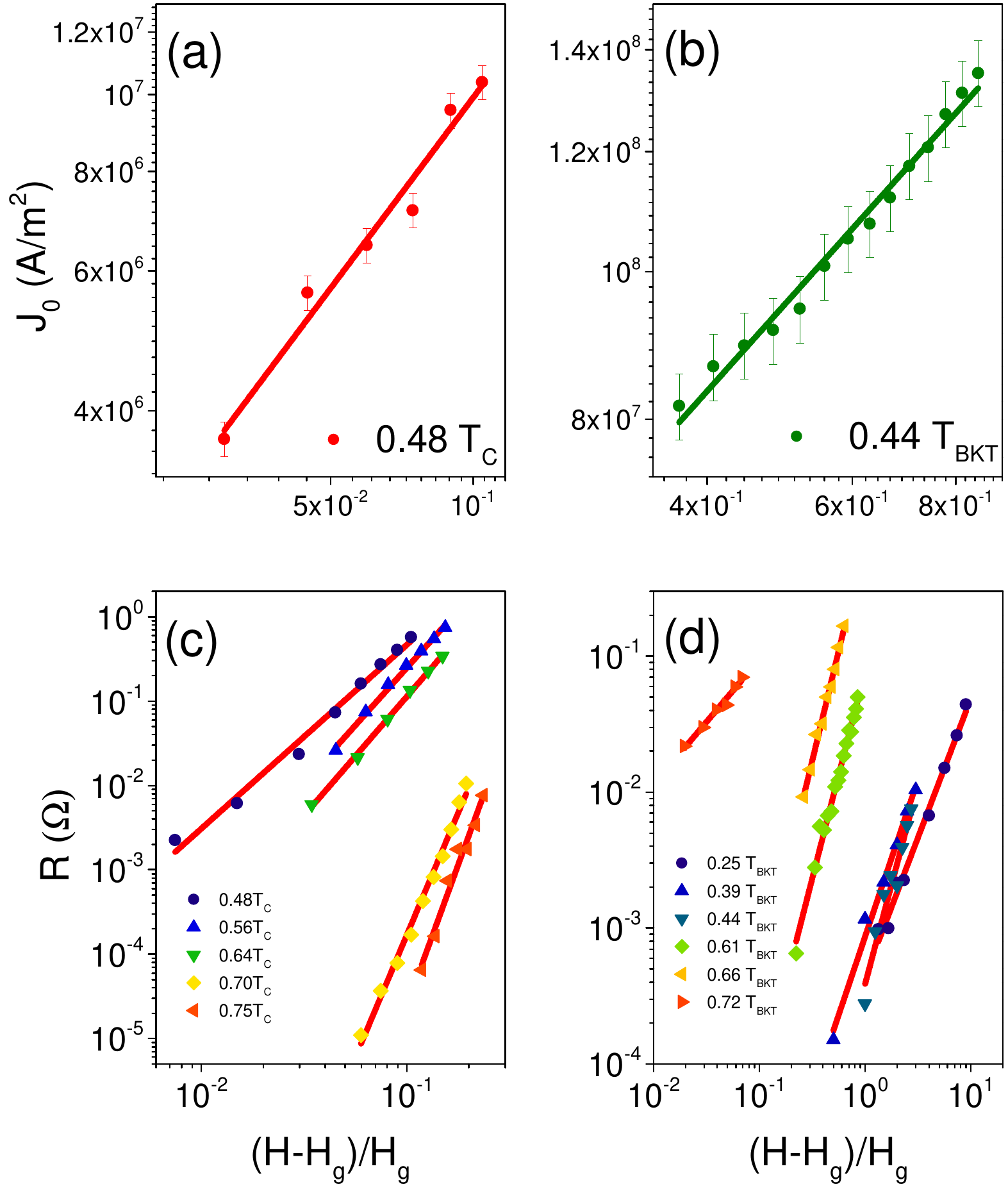}
		\small{\caption{(a) and (b) plots showing the variation of critical current density $J_0$ as a function of the scaled magnetic field $(H-H_g)/H_g$ for the 3D and 2D NbN film respectively in the superconducting regime  at one representative $T$. The solid lines are the power law fits (Eqn.~\ref{eqn:j0}) from which the  dynamical exponent $\alpha$ was extracted, as described in main text. Linear resistance, $R_L$ obtained from $E-J$ characteristics curves as a function of $(H-H_g)/H_g$ for different temperatures for the (c) 3D NbN superconductor, and (d) 2D NbN superconductor. The dashed red lines are the fits to Eqn.~\ref{bgmr} from which the quantity $\alpha(z+2-d)$ were extracted. The values of $\alpha(z+2-d)$ have been tabulated in the main text.
				\label{fig:J0_R}}}
	\end{center}
\end{figure}

\begin{table*}
	\caption{Critical exponents extracted from $E$-$J$ and magnetoresistance measurements  for  2D and 3D superconductors}
	\label{2d3dtable}
	\begin{tabular}{|c|c|c|c|c|c|c|c|c|c|}
		\hline
		\multicolumn{5}{|c|}{3D}                                                                                                                                                           & \multicolumn{5}{c|}{2D}                                                                                                                                                            \\ \hline
		T   & $\alpha$     & z    & \begin{tabular}[c]{@{}c@{}}$\alpha$(z+2-d) from\\ measured $\alpha$ and $z$\end{tabular} & \begin{tabular}[c]{@{}c@{}}$\alpha$(z+2-d) \\ extracted from  $\rho_L(H)$\end{tabular} & T    & $\alpha$    & z    & \begin{tabular}[c]{@{}c@{}}$\alpha$(z+2-d) from \\ measured $\alpha$ and $z$\end{tabular} & \begin{tabular}[c]{@{}c@{}}$\alpha$(z+2-d) \\ extracted from  $\rho_L(H)$\end{tabular} \\ \hline
		5.6 & 0.78 & 4.44 & 2.60                                                                         & 2.68                                                                           & 1.3 & 0.34 & 1.24 & 0.42                                                                         & 3.91                                                                           \\ \hline
		7.0   & 0.63  & 5.26 & 2.68                                                                         & 2.68                                                                           & 3.5  & 0.59 & 0.74 & 0.44                                                                         & 3.33                                                                           \\ \hline
		8.7 & 0.95  & 5.8  & 4.56                                                                         & 4.22                                                                           & 5.5  & 0.57 & 1.10 & 0.63                                                                         & 3.37                                                                           \\ \hline
		9.6 & 1.18 & 4.73 & 4.38                                                                         & 4.00                                                                           & 6.5  & 0.38 & 1.10 & 0.42                                                                         & 2.04                                                                          
\\ \hline
		10.0  & 0.99  & 5.82 & 4.77                                                                         & 4.15                                                                           &  8.0    & 0.68 & 0.68 & 0.46                                                                         & 1.45                                                                           \\ \hline
	\end{tabular}
\end{table*}

To verify the consistency of the scaling relations and exponents obtained near the transition from the $E$-$J$ plots, we looked into the plots of magnetic field dependence of linear resistivity, $\rho_L(H)$ versus ($H-H_g$) using Eqn.~\ref{bgmr}. The plot is shown in Fig.~\ref{fig:J0_R}(c) and (d) for 3D and 2D respectively and the exponents obtained are listed in table~\ref{2d3dtable}. 

We find an  excellent match in the values of the quantity $\alpha(z+2-d)$ obtained by the two independent methods -- from $E$-$J$ plots (using Eqn.~\ref{bgiv} and Eqn.~\ref{eqn:j0}) and from magnetoresistance plots (using Eqn.~\ref{bgmr})  for the 3D superconductor. But for the 2D superconductor, the values obtained from $E$-$J$ and magnetoresistance measurements differ by almost an order of magnitude. This mismatch in consistency of exponents for 2D superconductor can primarily be due to two reasons -- firstly this can be understood to indicate the absence of  vortex-solid/glass in 2D superconductors. Secondly, it is possible that the assumptions regarding a well defined dependence of the coherence length on $H$ ($\xi \sim \lvert H-H_g\rvert^{-\alpha}$) near the critical point and an exact scaling function  at critical point does not hold in the case of 2D.

\begin{figure}[t]
	\begin{center}
		\includegraphics[width=0.48\textwidth]{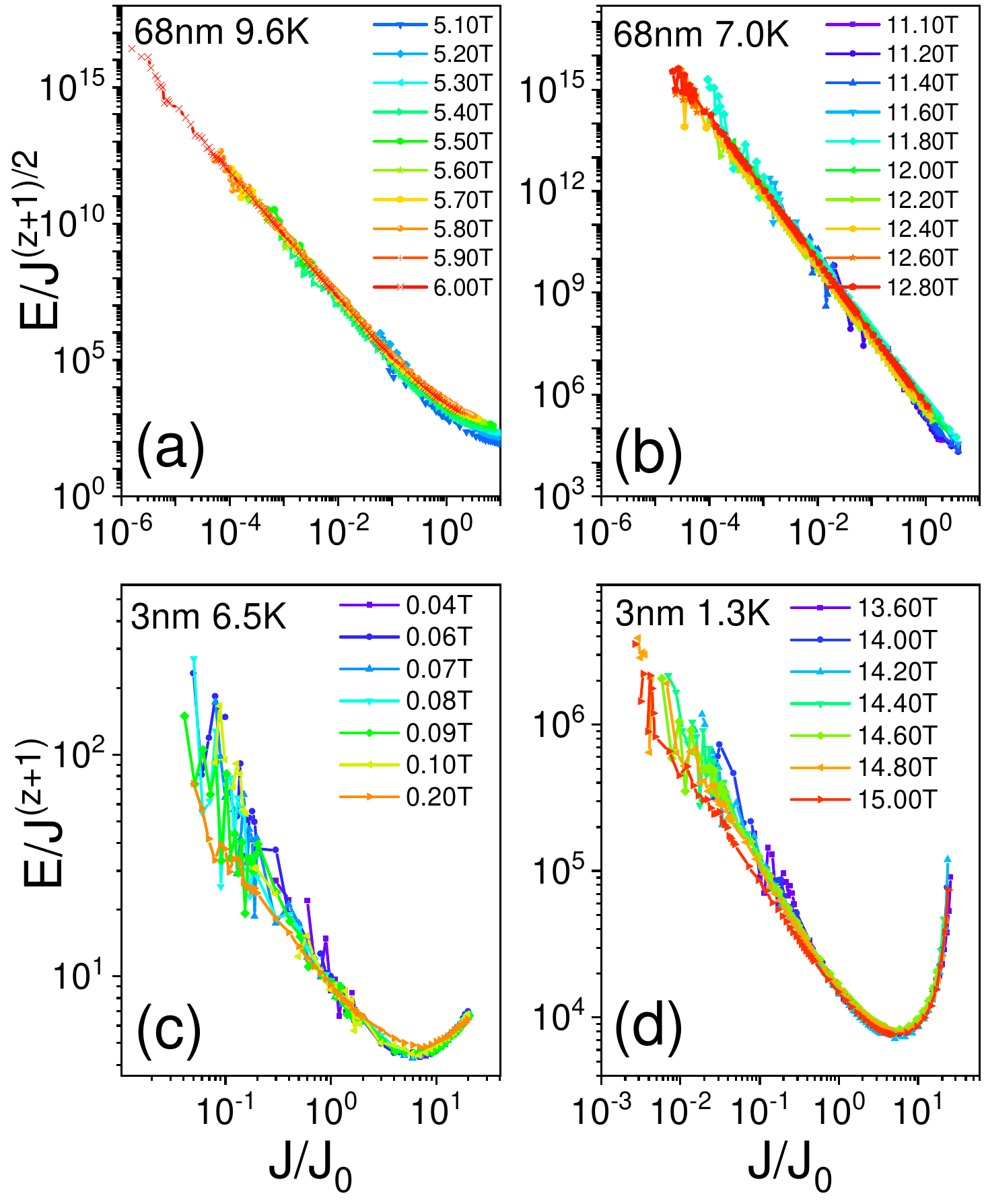}
		\small{\caption{(a) and (b): scaling plots of dc $E$-$J$ curves for 3D vortex-glass at $T$=9.6~K and 7.0~K, respectively. (c) and (d): scaling plots of dc $E$-$J$ curves for 2D vortex-glass at $T$=6.5~K and 1.3~K, respectively. The data in all cases show excellent agreement with the corresponding scaling relations [for details see text].
				\label{fig:Graph4}}}
	\end{center}
\end{figure}

To clarify this point, the $E$-$J$ curves obtained near the transition from vortex-solid/glass to a vortex-fluid were fitted to a scaling form derived from Eqn.~\ref{mastereq} which turned out to be 
\begin{equation}\label{master3deq}
\frac{E}{J^{(z+1)/2}}(\frac{K_BT}{\phi_0})^{(z-1)/2} \approx F_\pm(J/J_0)
\end{equation}
for the 3D superconductor, and 
\begin{equation}\label{master2deq}
\frac{E}{J^{(z+1)}}(\frac{K_BT}{\phi_0})^{z} \approx \acute{F_\pm} (J/J_0)
\end{equation}
for the 2D superconductor. The details of the derivation are given in Supplementary materials. The scaling relation indicates that  if one plots $E/J^{(z+1)\beta}$ ($\beta$ = 0.5 for 3D and 1 for 2D)as a function of ${J}/{J_0}$ for different magnetic fields at a fixed temperatures, the plots should collapse on top of each other. Fig.~\ref{fig:Graph4}(a) and (b) are the scaled plots of $E$ versus $J$  for 3D superconductor following Eqn.~\ref{master3deq} at two different $T$. The corresponding plots for the 2D superconductor are presented in Fig.~\ref{fig:Graph4}(c) and (d).  In all cases, we find the scaling relation to hold extremely well  as all the plots at a given $T$ fall on top of each other.  Similar analysis for T=8.7K and 6.0K for 3D and T=4K and 3.5K for 2D are presented in supplementary material. This excellent scaling agreement for the 2D case (as good as for the 3D superconductor) shows that  Eqn.~\ref{mastereq} holds for 2D superconductors. This suggests the possibility of the existence of vortex-solid/glass in 2D superconductors. This leaves us with the second possibility discussed earlier to explain the inconsistencies in the critical exponents extracted from $E$-$J$ and magnetoresistance measurements --  the assumed presence of a length scale relation of $\xi$ ($\sim \lvert H-H_g\rvert^{-\alpha}$) close to critical point  is probably not valid for 2D superconductors.

%\section{Conclusions}
To conclude, in this article we have looked for signatures of vortex dynamics in 3D and 2D superconductors  in NbN thin films. We have established a characteristic difference in response of the 2D  to 3D superconductors to a perpendicular magnetic field close to $T=T_C$. In 3D superconductor we find signatures of vortex-solid/glass to fluid transition. We extracted the  corresponding dynamical exponents from two independent measurements - magnetoresistance and current-voltage characteristics and found excellent match between the two sets.  On the other hand, for the  2D superconductor, the exponents obtained from these two disparate sets of experiments are  significantly different. We established that this anomaly  is due to the fact that the characteristic length scale diverges near the critical point in 2D superconductors in a manner different from that in the 3D superconductors. Our measurements and analysis indicates the existence of a vortex-solid/glass to fluid transition in 2D superconductors by validating the scaling form of fundamental $E$-$J$ relation near the critical point both in the case 2D as well as for 3D superconductors (where such transitions are well documented).

%\section{Acknowledgements}
AB acknowledges funding from Nanomission and FIST program, Department of Science \& Technology (DST), Government of India. HKK acknowledges CSIR, Govt of India and IISc, Bangalore for financial help.

%\bibliography{NbN_2D3D}


\newpage	
\begin{center}

\section*{Supplementary Information}

%\section*{Supplementary Information:\\Universal scaling behavior near vortex-glass/solid to vortex-fluid transition in type-II superconductors in h two- and three-dimensions}

\author{Hemanta Kumar Kundu}
\affiliation{Department of Physics, Indian Institute of Science, Bangalore 560012, India}
\author{John Jesudasan}
\affiliation{Tata Institute of Fundamental Research, Mumbai 400005, India}
\author{Pratap Raychaudhuri}
\affiliation{Tata Institute of Fundamental Research, Mumbai 400005, India}
\author{Subroto Mukerjee}
\affiliation{Department of Physics, Indian Institute of Science, Bangalore 560012, India}
\author{Aveek Bid}
\email{aveek@iisc.ac.in}
\affiliation{Department of Physics, Indian Institute of Science, Bangalore 560012, India}

\maketitle

\end{center}

	%\correspondingauthor{\textsuperscript{1}To whom correspondence should be addressed. E-mail: aveek.bid@physics.iisc.ernet.in}

%\section*{\large{Supplementary Information}}
\renewcommand{\thefigure}{S\arabic{figure}}
\setcounter{figure}{0}

\section{$E$-$J$ curves for 3D and 2D superconductor:} 
\begin{figure}[h]
	\begin{center}
		\includegraphics[width=0.5\textwidth]{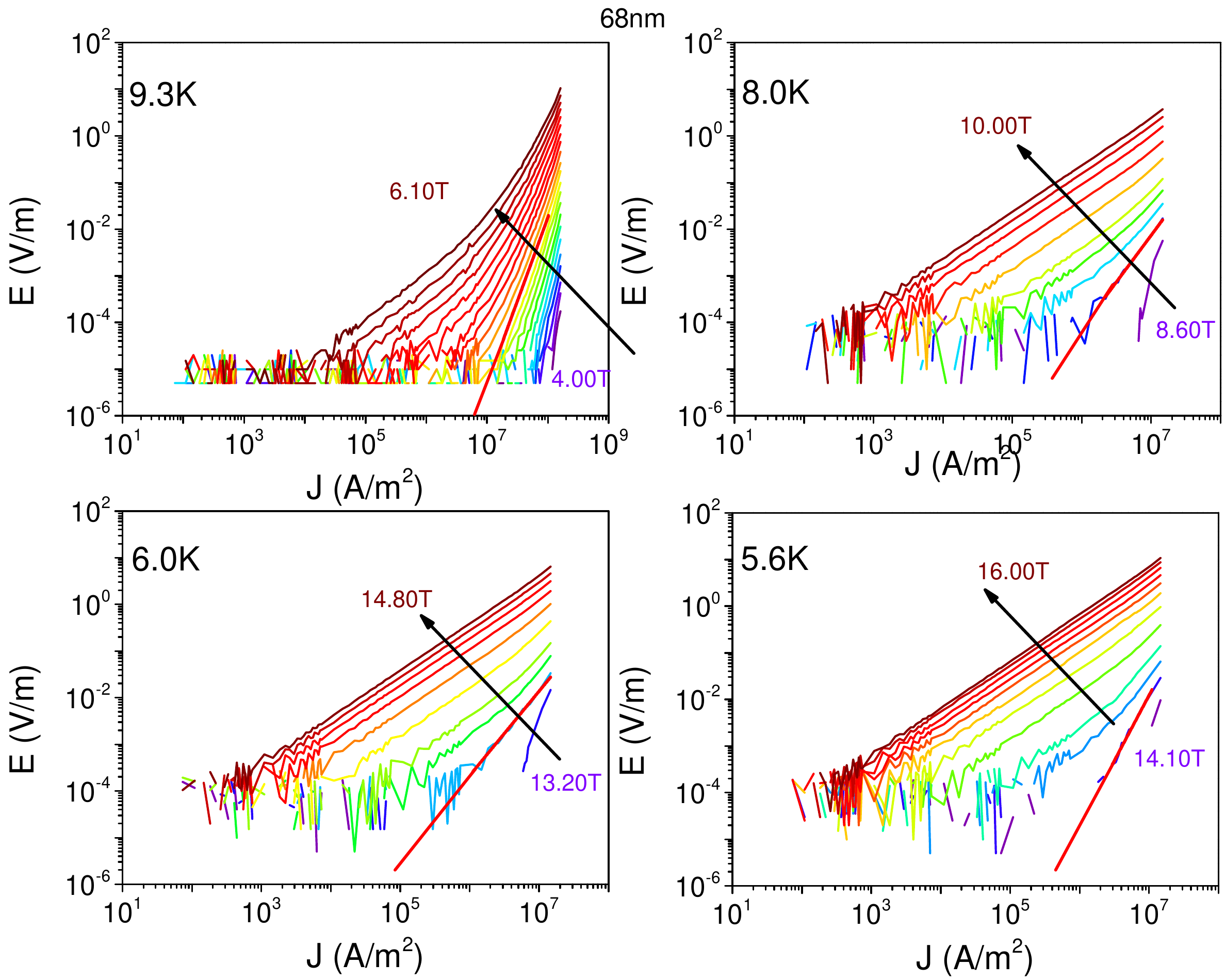}
		\small{\caption{\textbf{vortex-solid/glass melting in 3D NbN superconductor:} $E-J$ characteristics of 3D (68~nm) NbN superconductor in the presence of $H$ at constant temperatures $T$ = 9.3~K, 8.0~K, 6.0~K and 5.6~K. The red thick lines are the power-law fit to the curve at $H$ corresponding to the melting field, $H_g$. 
				\label{fig:EJ_68nm}}}
	\end{center}
\end{figure}

\begin{figure}[h]
	\begin{center}
		\includegraphics[width=0.5\textwidth]{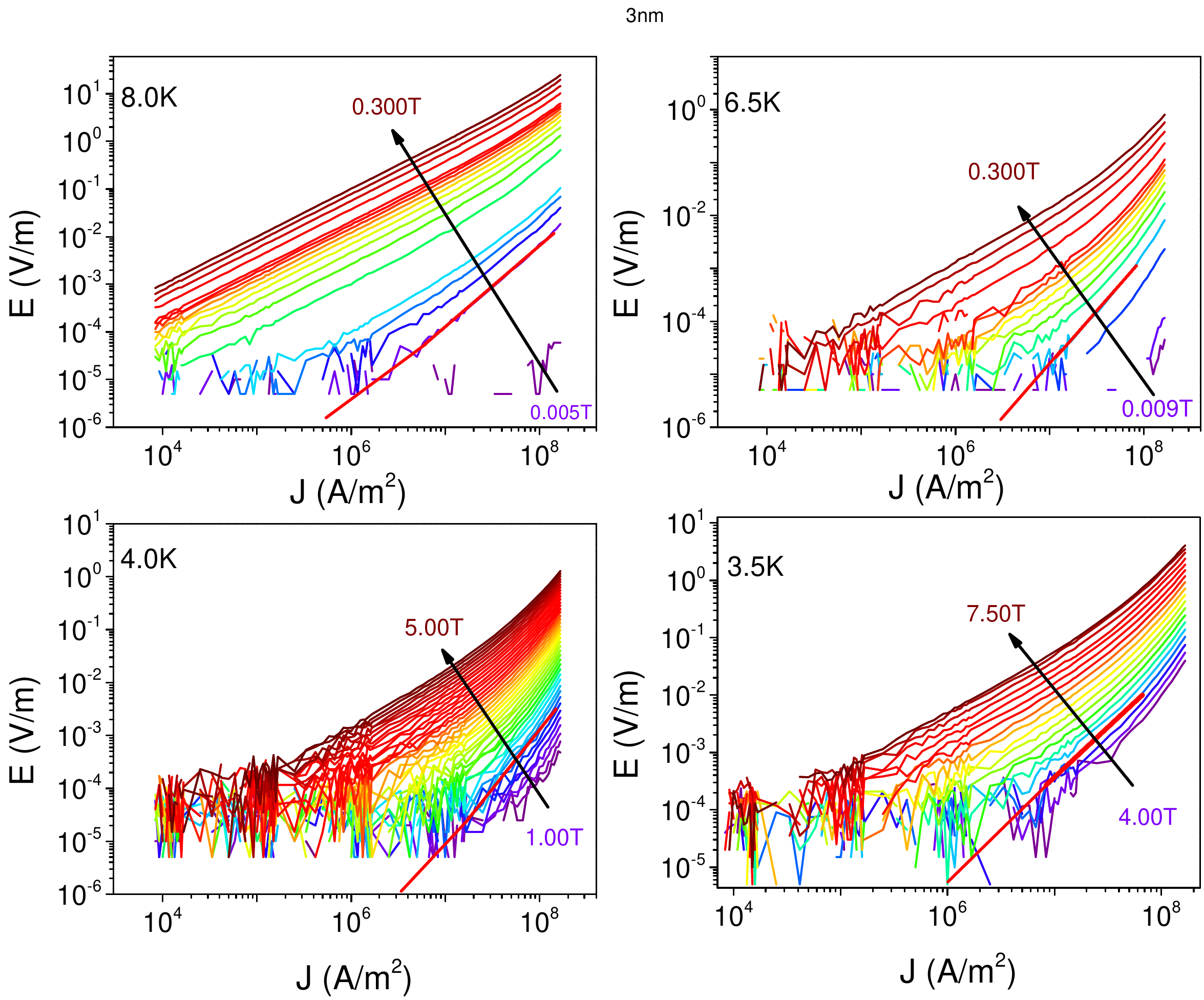}
		\small{\caption{\textbf{vortex-solid/glass melting in 2D NbN superconductor:} $E-J$ characteristics of  2D (3~nm) NbN superconductor in the presence of $H$ at constant temperatures $T$ = 8.0~K, 6.5~K, 4.0~K and 3.5~K. The red thick lines are the power-law fit of the curve at $H$ corresponding to the melting field, $H_g$. 
				\label{fig:EJ_3nm}}}
	\end{center}
\end{figure}

Figure~\ref{fig:EJ_68nm} (a), (b), (c) and (d) are the $E$-$J$ plots for $T$ = 9.3~K, 8.0~K, 6.0~K and 5.6~K respectively for 3D (68nm) NbN superconductor. For 2D (3nm) NbN superconductor, the $E$-$J$ curves for T = 8.0~K, 6.5~K, 4.0~K and 3.5~K are shown in fig.~\ref{fig:EJ_3nm}(a), (b), (c), (d)  respectively. At every temperature, for both 2D and 3D superconductors, at a specific value of applied magnetic field, which we identify to be $H_g$, we observe a power law dependence of $E$ on $J$,  as explained in main text.  The power behaviour was shown to follow the relation: 
$$E(J;H=H_g) \approx J^{(z+1)/(d-1)}$$
The red thick line shows the fits. From the fits, the values of the dynamical exponent $z$ at different $T$ were  extracted and have been presented in the main text. 

\section{Scaling behavior of $E-J$ curves:} 
\begin{figure}[h]
	\begin{center}
		\includegraphics[width=0.5\textwidth]{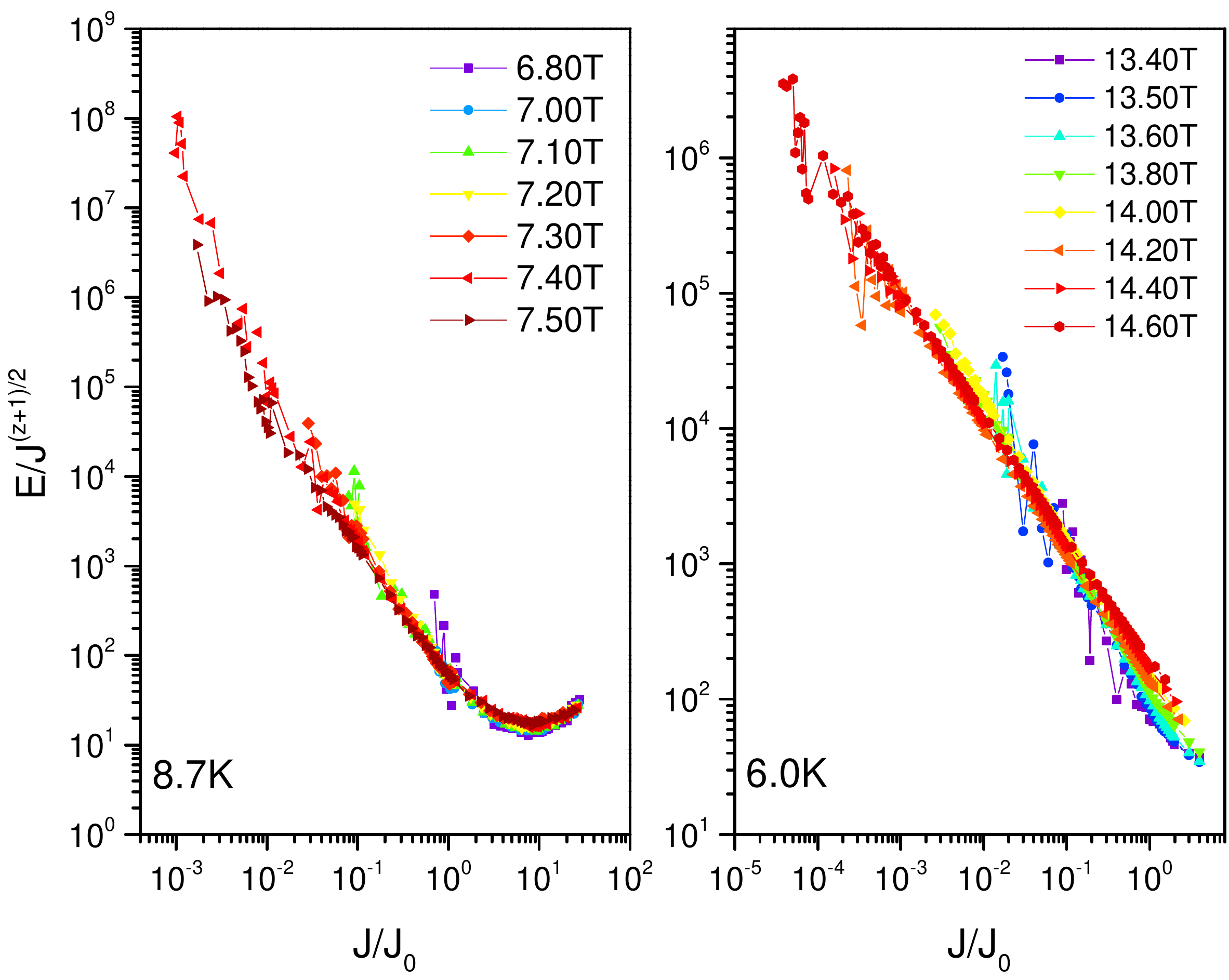}
		\includegraphics[width=0.5\textwidth]{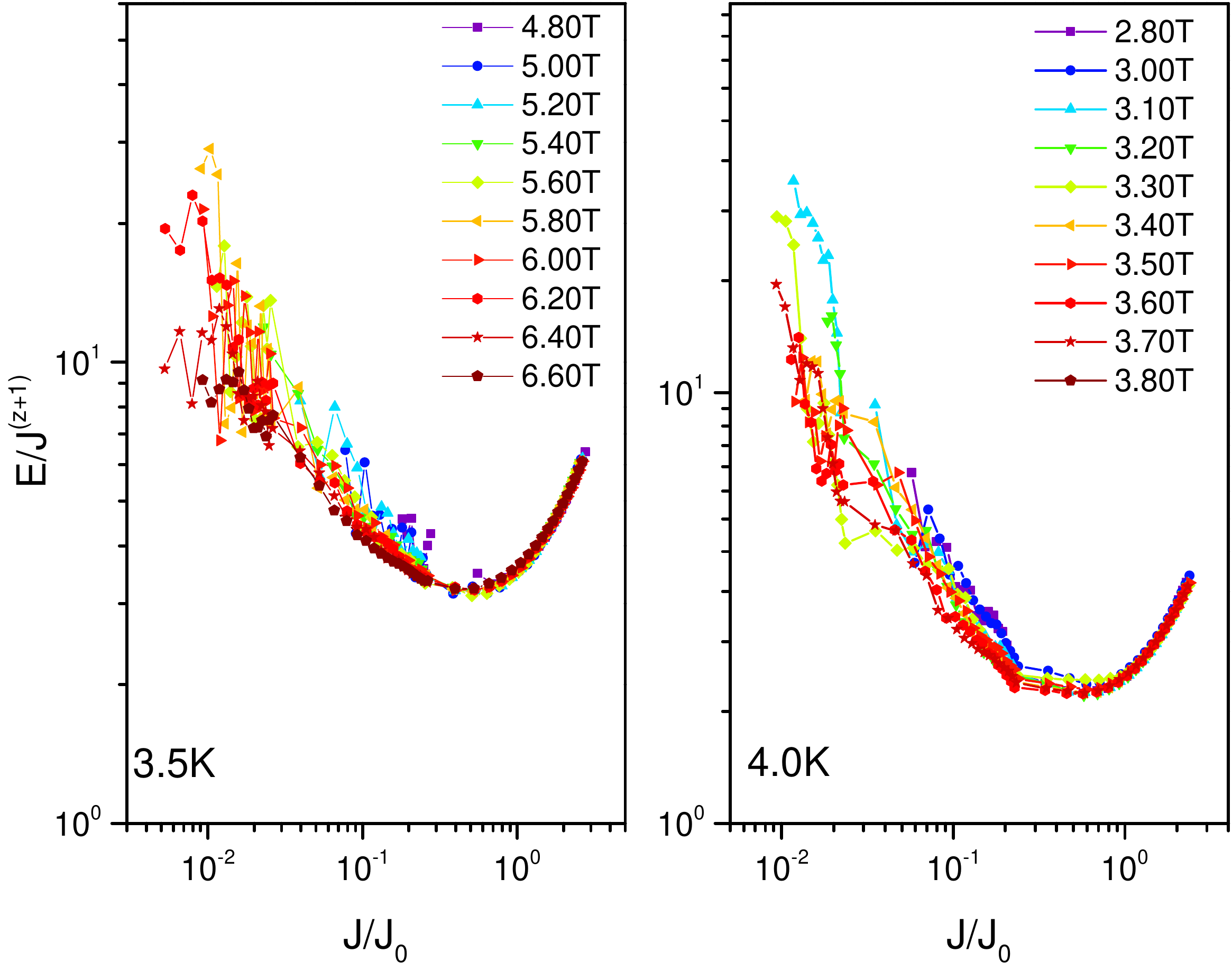}
		\small{\caption{\textbf{Scaling of $E-J$ curves close to vortex-solid/glass to vortex-fluid transition} Scaling plots of dc $E-J$ curves (following Eqn.~\ref{Eqn:3dscaling}) obtained at (a) T = 8.7~K and (b) 6.0~K for the 3D superconductor. Corresponding scaling plots  of dc $E-J$ curves (following Eqn.~\ref{Eqn:2dscaling}) obtained at (c) T = 3.5~K and (d) 4.0~K  for the 2D superconductor.
				\label{fig:scaling3D}}}
	\end{center}
\end{figure}

The fundamental E-J characteristics for vortex glass transition in d-dimensional superconductor is given by,
\begin{equation}\label{masterIV}
E(J) \approx J\xi^{d-2-z}\tilde{E}_\pm(J\xi^{d-1}\phi_0/K_BT)
\end{equation}
As $J_0 = \frac{ K_B T}{\phi_0 \xi ^{d-1}}$, 
\begin{equation}\label{IV_d} 
E(J) \approx J\bigg( \frac{ K_B T}{\phi_0 J_0}\bigg)^\frac{d-2-z}{d-1}\tilde{E}_\pm(J/J_0)
\end{equation}
For d=3,
$$E(J) \approx J\bigg( \frac{ K_B T}{\phi_0 J_0}\bigg)^\frac{1-z}{2}\tilde{E}_\pm(J/J_0)$$of dc $E-J$ curves (following Eqn.~\ref{Eqn:2dscaling}) obtained at T = 3.5~K and 4.0~K  for the 2D superconductor.
$$\frac{E(J)}{J^\frac{1+z}{2}} \approx \bigg( \frac{ K_B T}{\phi_0}\bigg)^\frac{1-z}{2}\bigg(\frac{J}{J_0}\bigg)^\frac{1-z}{2} \tilde{E}_\pm(J/J_0)$$

This yields the following scaling relation for 3D superconductor near the vortex solid/glass to vortex fluid melting transition:
\begin{eqnarray}
\frac{E}{J^{(z+1)/2}}\bigg(\frac{K_BT}{\phi_0}\bigg)^{\frac{z-1}{2}} \approx F_\pm(J/J_0)
\label{Eqn:3dscaling}
\end{eqnarray}

%\begin{figure}[t]
%	\begin{center}
%		\includegraphics[width=0.5\textwidth]{scaling2D.pdf}
%		\small{\caption{Scaling plots of dc $E-J$ curves (following Eqn.~\ref{Eqn:2dscaling}) obtained at T = 3.5~K and 4.0~K  for the 2D superconductor.
%				\label{fig:scaling2D}}}
%	\end{center}
%\end{figure}
In Eqn.\ref{IV_d} if we use $d$=2,  we  obtain the corresponding scaling form  for the 2D superconductor in the vicinity of the vortex solid/glass to vortex fluid melting transition:
\begin{eqnarray}
\frac{E}{J^{(z+1)}}\bigg(\frac{K_BT}{\phi_0}\bigg)^{z} \approx \acute{F_\pm} (J/J_0)
\label{Eqn:2dscaling}
\end{eqnarray}
Plots of the scaling relation Eqn.~\ref{Eqn:3dscaling} for the  3D superconductor at two different temperatures $T$= 6.0~K and 8.7~K have been shown in fig.~\ref{fig:scaling3D}(a) and (b) respectively. Similar scaling plots of Eqn.~\ref{Eqn:2dscaling}  for the 2D superconductor at $T$ = 3.5~K and 4.0~K are shown in Fig.~\ref{Eqn:3dscaling}(c) and (d) respectively. At all temperatures, for both the 2D and 3D superconductors, the $E$-$J$ data obtained for different values of magnetic field close to $H=H_g$ scale extremely well.

\end{document}